\documentclass{SCGE}

\usepackage{multicol}
\usepackage{cite}
\usepackage[dvips]{graphicx}
\usepackage{color}

\usepackage[colorlinks,%
            citecolor=blue,linkcolor=blue,hypertex, %
            breaklinks=true]{hyperref}

\begin{document}


\def\bm{\boldsymbol}

\Year{2011} %
\Month{??}
\Vol{41} %
\No{??} %
\BeginPage{1} %
\EndPage{5} %
\AuthorMark{{\rm C. J. XIA,} et al.}
\DOI{} 

\title[Systematic Study of Survival Probability of Excited SHN]%
      {Systematic Study of Survival Probability of Excited Superheavy Nuclei}%

\author[1]{Cheng-Jun XIA}{}
\author[1]{Bao-Xi SUN}{}
\author[2,3]{En-Guang ZHAO}{}
\author[2,3]{Shan-Gui ZHOU}{Email: sgzhou@itp.ac.cn}

\address[{\rm1}]{College of Applied Sciences,
                 Beijing University of Technology, Beijing 100124, China;}
\address[{\rm2}]{Institute of Theoretical Physics, Chinese Academy of Sciences,
                 Beijing 100190, China;}
\address[{\rm3}]{Center of Theoretical Nuclear Physics, National Laboratory
                 of Heavy Ion Accelerator, Lanzhou 730000, China}

\maketitle \vspace{-3.5mm} {\footnotesize
 \begin{center}
  Received November 23, 2010; accepted December 20, 2010
 \end{center}
}%
\vspace*{-5mm}

\begin{center}
\rule{16.5cm}{0.4pt}
\parbox{16.5cm}
{
\begin{abstract}
The stability of excited superheavy nuclei (SHN) with $100 \leq Z
\leq 134$ against neutron emission and fission is investigated by
using a statistical model. In particular, a systematic study of the
survival probability against fission in the 1n-channel of these SHN
is made. In present calculations the neutron separation energies and
shell correction energies are consistently taken from the calculated
results of the finite range droplet model which predicts an island of
stability of SHN around $Z=115$ and $N=179$. It turns out that this
island of stability persists for excited SHN in the sense that the
calculated survival probabilities in the 1n-channel of excited SHN at
the optimal excitation energy are maximized around $Z=115$ and
$N=179$. This indicates that the survival probability in the
1n-channel is mainly determined by the nuclear shell effects.
\end{abstract}
}
\end{center}
\vspace*{-0.6cm}
\begin{center}
\parbox{16.5cm}
{\bf\jiuhao survival probability, superheavy elements, island of
stability, neutron emission, fission}
\end{center}

\begin{center}
\parbox{16.5cm}{\PACS{\hspace*{-2mm}\rm 24.10.Pa, 24.60.Dr, 25.85.-w, 27.90.+b, 21.10.-k}
\rule{16.5cm}{0.4pt}}
\end{center}

\wuhao\vspace*{1.5mm}
\renewcommand{\baselinestretch}{1.08} \baselineskip 12.2pt\parindent=10.8pt

\no 

\section{Introduction~\label{sec:intro}}

The importance of quantum shell effects in stabilizing heavy nuclei
was realized and the existence of superheavy elements was predicted
in 1960s~\cite{Myers1966, Sobiczewski1966, Meldner1967}.
Since then, a lot of efforts has been made to explore the
island of stability of superheavy nuclei (SHN). In experiment,
via cold fusion reactions, superheavy elements with $102 \le Z
\le 112$ have been synthesized in GSI~\cite{Hofmann2000} and
that with $Z = 113$ in RIKEN~\cite{Morita2004}. Superheavy
elements with $Z$ up to 118 have also been synthesized with hot
fusion reactions in Dubna~\cite{Oganessian2007,
Oganessian2010}. Theoretical investigations of SHN are focused
both on the structure and decay
properties~\cite{Lalazissis1996, Rutz1997, Long2002, Ren2003,
Zhang2005, Zhang2006} and on the synthesis
mechanism~\cite{Li2003, Zhang2006c, Feng2007, Shen2008,
Liu2009a, Zagrebaev2010}.

A superheavy nucleus in its ground state can decay via
spontaneous fission, $\alpha$-decay, $\beta$-decay, etc.. As a
crucial ingredient of the existence of the island of stability,
the stability of superheavy nuclei in their ground states has
been studied extensively since the existence of superheavy
elements was predicted~\cite{Nilsson1969}. The stability of an
excited superheavy nuclei is also an important issue. On the
one hand, it helps us in understanding the stability behavior
of a superheavy nucleus against excitation, e.g., the
attenuation of the shell effects with temperature. On the other
hand, the survival probability $W_\mathrm{sur}$ of an excited
compound nucleus against processes in which the charge number
is changed is directly related with the stability of an excited
superheavy nucleus in various channels. The survival
probability $W_\mathrm{sur}$ is an important factor in the the
study of synthesis mechanism of superheavy elements. Presently
most of the calculations are focused on the stability of
superheavy compound nuclei formed in cold and/or hot fusion
reactions~\cite{Adamian2000, Zubov2002, Zubov2005, Zubov2009}.

In order to study the influence of the shell effects on the
stability of superheavy nuclei with excitation, we carry out a
systematic investigation of the stability of excited superheavy
nuclei. In the present work, the stability of excited
superheavy nuclei with $100 \leq Z \leq 134$ against neutron
emission and fission is studied by using a statistical model.
As an example, we present a systematic study of the survival
probability against fission in the 1n-channel of these
superheavy nuclei. It is well known that there are quite
different predictions of the shell structure of superheavy
nuclei from different models. As a consequence, the next proton
magic number after 82 could be 114, 120, 126, etc.. In this
paper, the properties of superheavy nuclei, e.g., the neutron
separation energies and shell correction energies, are
consistently taken from predictions of the finite range droplet
model~\cite{Moeller1995}.

The paper is organized as follows. In Sec.~\ref{sec:theory}, the
formalism of the statistical model is briefly sketched. The results
and discussions are presented in Sec.~\ref{sec:results}. Finally a
summary is given in Sec.~\ref{sec:summary}.

\section{Formalism~\label{sec:theory}}

An excited superheavy compound nucleus can decay via fission or
emitting photon(s), neutron(s), proton(s) or light-charged
particle(s) like $\alpha$-particle. Among all these channels, the
most favorable ones are fission and neutron(s) emission. In the
present study, we mainly focus on these two channels.

The decay width for the neutron emission of a compound nucleus
with excitation energy $E^*$ and spin $J$ is calculated as
\cite{Feng2006}
\begin{equation}
 \Gamma_\text{n}(E^*,J) =
 \frac{2 m_\text{n} R^2}{\pi\hbar^2}
 \int_0^{E^*-S_\mathrm{n}}
  \frac{ \rho(E^*-S_\mathrm{n}-\varepsilon_\mathrm{n},J) }
       { \rho(E^*,J) }
  \varepsilon_\mathrm{n}
  d\varepsilon_\mathrm{n}
 \ ,
 \label{eq:width_n}
\end{equation}
where $m_\mathrm{n}$ is the neutron mass, $S_\mathrm{n}$ is the
neutron separation energy, $R=r_0 A^{1/3}$ is the radius of the
compound nucleus with $r_0=1.20$ fm, and $\rho(E^*,J)$ is the
level density which is discussed later.

The fission width can be calculated with the Bohr-Wheeler
formula as \cite{Bohr1939}
\begin{equation}
 \Gamma_\mathrm{f}(E^*,J)
 =
 \frac{ 1 }{ 2\pi }
 \int_0^{E^*-B_\mathrm{f}}
  \frac{ \rho_\mathrm{s.d.}(E^*-B_\mathrm{f}-\varepsilon_\mathrm{f},J) }
       { \rho_\mathrm{s.d.}(E^*,J) }
  T_\mathrm{f}(\varepsilon_\mathrm{f})
  d\varepsilon_\mathrm{f}
 \ ,
 \label{eq:width_f}
\end{equation}
where $B_\mathrm{f}$ is the fission barrier,
$\rho_\mathrm{s.d.}(E^*,J)$ is the level density at the saddle
point, and $T_\mathrm{f}(\varepsilon_\mathrm{f})$ is the
barrier transmission probability,
\begin{equation}
 T_\mathrm{f}(\varepsilon_\mathrm{f})
 =
 \left\{ 1 + \exp\left[- \frac{2\pi \varepsilon_\mathrm{f}}{\hbar\omega_\mathrm{s.d.}}
                 \right]
 \right\}^{-1}
 \ ,
\end{equation}
with the barrier width $\hbar\omega_\mathrm{s.d.}=2.2\
\mathrm{MeV}$~\cite{Zubov2005}. The fission barrier height
including the washing out effect of shell effects with the
excitation energy $E^*$ and spin $J$ is given as \cite{Li2004a}
\begin{eqnarray}
 B_\mathrm{f}(E^*, J)
 & = &
 B_\mathrm{f}^\mathrm{Mac}
 +
 B_\mathrm{f}^\mathrm{Mic} \exp\left( -\frac{E^*}{E_\mathrm{D}} \right)
 -
 \left( \frac{ \hbar^2 }{ 2{\cal J}_\mathrm{g.s.} }
      - \frac{ \hbar^2 }{ 2{\cal J}_\mathrm{s.d.} }
 \right)
 J(J+1)
 \ ,
 \label{eq:Bf}
\end{eqnarray}
where $B_\mathrm{f}^\mathrm{Mac}$ is the macroscopic part of
the fission barrier height of the compound
nucleus~\cite{Dahlinger1982}. Under the assumption that the
shell correction energy at the saddle point is negligible, the
microscopic part of the fission barrier height
$B_\mathrm{f}^\mathrm{Mic} = - E_\mathrm{Mic}$ with
$E_\mathrm{Mic}$ the shell correction energy in the ground
state. There are several expressions for the damping parameter
$E_\mathrm{D}$~\cite{Ignatyuk1975, Schmidt1991} and we take
$E_\mathrm{D} = 5.48A^{1/3} / ( 1+1.3A^{-1/3}
)$~\cite{Ignatyuk1975}. The moments of inertia of the compound
nucleus in its ground state and at the saddle point are
calculated as
\begin{equation}
 {\cal J}_\mathrm{g.s.(s.d.)}
 =
 \frac{2}{5} M R^2 \left( 1 + \beta_2^\mathrm{g.s.(s.d.)}/3 \right)
 \ .
\end{equation}

The level density is calculated from the Fermi-gas model
according to Ref.~\cite{Ignatyuk1979},
\begin{eqnarray}
 \rho(E^*,J)
 & = &
 \frac{ 2J+1 }{ 24 \sqrt{2} \sigma^3 a^{1/4} (E^*-\delta)^{5/4} }
 \exp\left[2\sqrt{a(E^*-\delta)}-\frac{(J+1/2)^2}{2\sigma^2}\right]
 \ ,
\end{eqnarray}
with
\begin{equation}
 \sigma^2
 =
 6 \bar{m}^2 \sqrt{a(E^\ast-\delta)}/\pi^2
 \ , \ \
 \bar{m}^2 \approx 0.24A^{2/3}
 \ ,
\end{equation}
where the level density parameter $a_\mathrm{s.d.} =
{1.1A}/{12}$ {MeV}$^{-1}$ at the saddle point and $a =
{A}/{12}$ {MeV}$^{-1}$ in other cases. The pairing correction
$\delta = 12 / \sqrt{A}$ MeV, 0 and $-12/\sqrt{A}$ MeV for
even-even, odd-even and odd-odd nuclei,
respectively~\cite{Zubov2009}.

The survival probability in the 1n-channel is calculated by
\begin{equation}
 W_\mathrm{sur}(E^*,J)
 =
 P_\mathrm{1n}(E^*,J)
 \frac{ \Gamma_\mathrm{n}(E^*,J) }
      { \Gamma_\mathrm{n}(E^*,J) + \Gamma_\mathrm{f}(E^*,J) }
 \ ,
 \label{eq:Wsur}
\end{equation}
where the realization probability for 1n-emission
reads~\cite{Adamian2000}
\begin{equation}
 P_\mathrm{1n}(E^*,J) = \exp\left[ -\frac{(E^*-S_n-2T)^2}{2\Sigma^2} \right]
 \ .
\end{equation}
The nuclear temperature $T = \left[{ 1 + \sqrt{1+4aE^*} }) \right] /
{2a}$, $\Sigma=2.2$ MeV and $a$ is the level density parameter.

\section{Results and discussions~\label{sec:results}}

For the superheavy nuclei, there are different predictions of
ground state and saddle point properties. Since both the
neutron emission and fission processes are connected closely to
the shell structure, in order to study the influence of the
shell effects on the stability of superheavy nuclei with
excitation, one should take the nuclear property parameters for
calculating the neutron emission and fission consistently from
one single model. In the present work, the properties of
superheavy nuclei with $100 \leq Z \leq 134$ are taken from
predictions of the finite range droplet
model~\cite{Moeller1995}.

\begin{figure}
\begin{center}
\includegraphics[width=16cm]{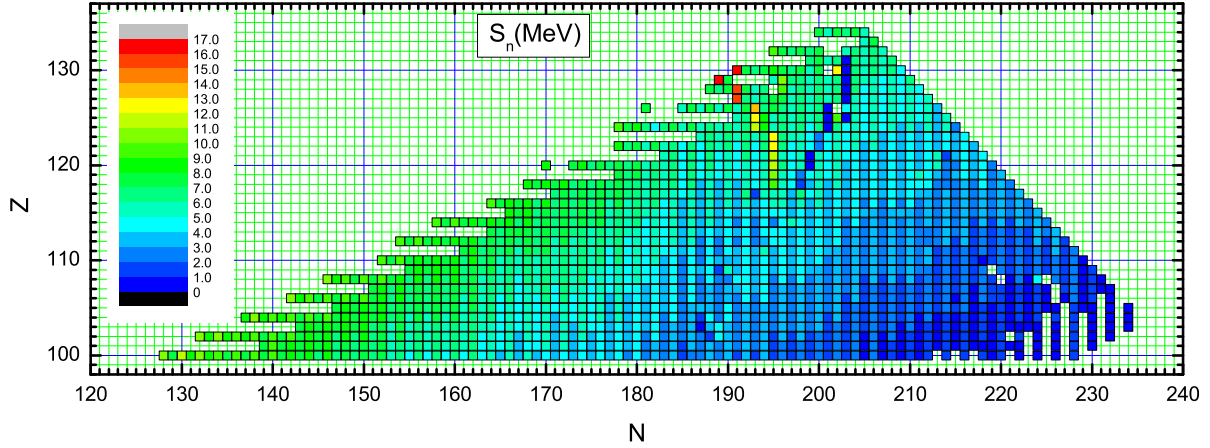}
\end{center}
\caption{\label{fig:Sn_Moeller1995}
The neutron separation energies of superheavy nuclei with $100 \leq Z
\leq 134$ predicted by the finite range droplet model~\protect\cite{Moeller1995}.}
\end{figure}

One of the crucial parameters for calculating the width of
neutron emission is the neutron separation energy.
Figure~\ref{fig:Sn_Moeller1995} shows the neutron separation
energies of superheavy nuclei with $100 \leq Z \leq 134$ from
the finite range droplet model~\cite{Moeller1995}. Only those
nuclei within the proton drip line and the neutron drip line
are included. One finds several common features in
Fig.~\ref{fig:Sn_Moeller1995}. First, the neutron separation
energy decreases with neutron number. Second, there is clearly
odd-even effects in the neutron separation energy. With these
values for the neutron separation energies and level density
parameters given in Sec.~\ref{sec:theory}, the width of neutron
emission is calculated from Eq.~(\ref{eq:width_n}).

\begin{figure}
\begin{center}
\includegraphics[width=16cm]{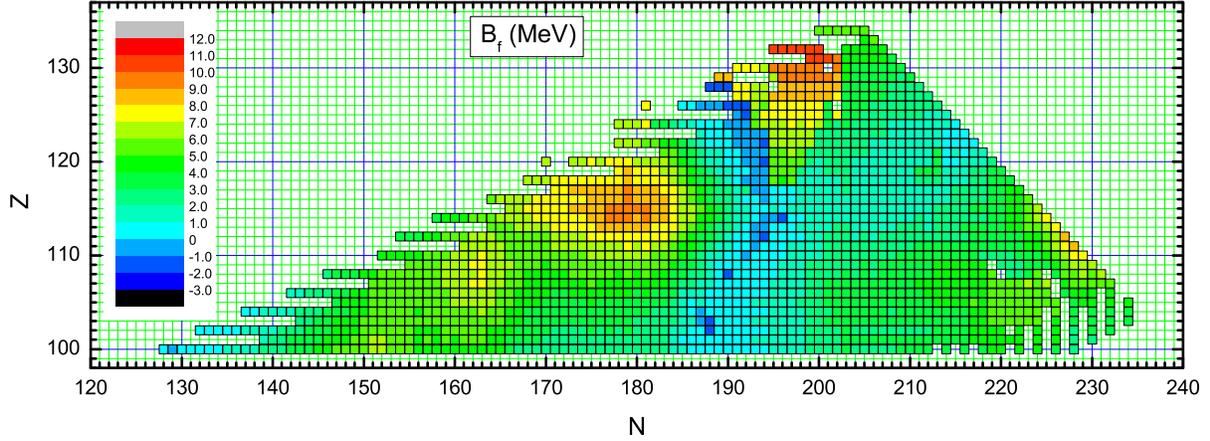}
\end{center}
\caption{\label{fig:Bf_Moeller1995}
The fission barriers of superheavy nuclei with $100 \leq Z
\leq 134$ at $E^*=0$ and $J=0$ calculated from Eq.~(\protect\ref{eq:Bf}) where
the shell correction energies are taken from the finite range droplet
model~\protect\cite{Moeller1995}.}
\end{figure}

The fission barrier heights of superheavy nuclei with $100 \leq
Z \leq 134$ at $E^*=0$ and $J=0$ are shown in
Fig.~\ref{fig:Bf_Moeller1995} with the shell correction
energies taken from the finite range droplet
model~\protect\cite{Moeller1995}. The fission barrier height of
a superheavy nucleus is mainly determined by the shell
correction energy because the macroscopic part $B^\mathrm{Mac}$
is quite small. According to Eq.~(\ref{eq:Bf}), a larger
negative shell correction energy results in a higher fission
barrier. The finite range droplet model predicts that the
island of stability centers around $Z=115$ and $N=179$ due to
the quantum shell effects. Correspondingly, there is a region
with higher fission barrier in Fig.~\ref{fig:Bf_Moeller1995}
around $Z=115$ and $N=179$. Meanwhile the deformed sub-shells
at $Z=108$ and $N=162 \sim 164$ also manifest themselves with
fission barriers as high as about $7 \sim 8$ MeV. In addition,
there is another mass region around $Z=130$ and $N=198$ in
which the superheavy nuclei also have very high fission
barriers. There is a roughly vertical band around $N=190$ in
which the fission barrier heights of the nuclei are negative.
This means that these nuclei do not exist according to the
finite range droplet model.

The survival probability of an excited superheavy nucleus with $J=0$
in the 1n-channel calculated from Eq.~(\ref{eq:Wsur}), as a function
of the excitation energy $E^*$, shows an anti-parabolic shape. We
take the maximal value in the $W_\mathrm{sur} \sim E^*$ curve for
each superheavy nucleus and define the corresponding $E^*$ as the
optimal excitation energy.

\begin{figure}
\begin{center}
\includegraphics[width=16cm]{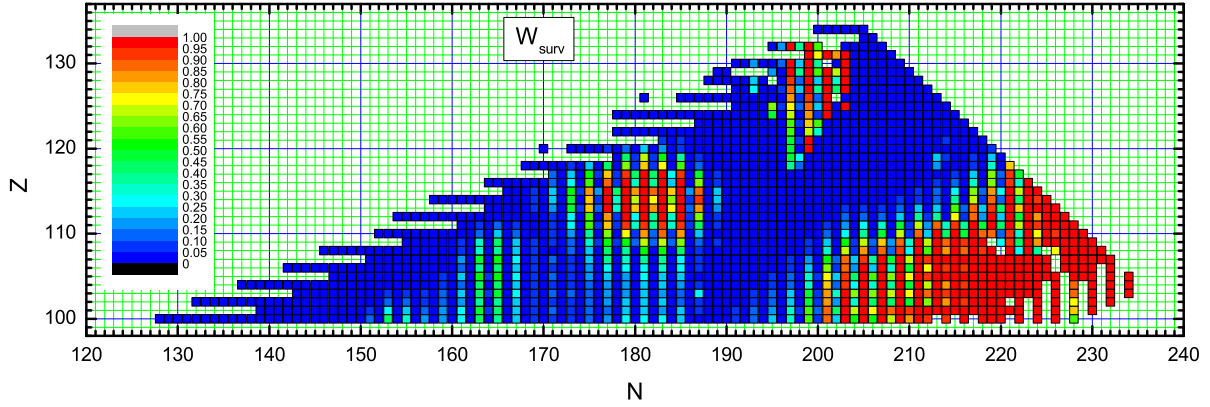}
\end{center}
\caption{\label{fig:Wsur}
The survival probability against fission of excited superheavy
nuclei in the 1n-channel with $100 \leq Z \leq 134$ at the optimal
excitation energy and $J=0$.}
\end{figure}

In Fig.~\ref{fig:Wsur} the survival probability of excited superheavy
nuclei with $100 \leq Z \leq 134$ at the optimal excitation energies
are given. There are two islands with larger survival probability in
Fig.~\ref{fig:Wsur} which roughly correspond to the islands of
superheavy nuclei with higher fission barriers as shown in
Fig.~\ref{fig:Bf_Moeller1995}. This indicates that for a superheavy
nucleus in these two mass regions, the fission width is quite small
due to the high fission barrier. In the very neutron-rich region, the
neutron separation energy is small which results in a large neutron
emission width. Therefore the survival probability in the 1n-channel
becomes larger in the very neutron-rich region if one only includes
the fission and neutron emission processes. For these nuclei, other
decay channels should be taken into account. The survival probability
of superheavy nuclei in the 1n-channel shows clear odd-even effects
which is mainly from the odd-even effects in the neutron separation
energy.

From the above discussions, one can conclude that the shell
effects also plays important roles in the stability behavior
and survival probability of excited superheavy nuclei.
Therefore nuclear parameters such as the separation energy and
fission barriers from models which predict different shell
structures in the region of superheavy nuclei may give very
different survival probability. This may also result in
different trends in the evaporation residue cross section as a
function of the proton number in the experimentally unknown
mass region.

Besides the neutron separation energy and the fission barrier which
are directly related to the shell effects, there are some other
parameters which are also very important in the calculation of the
survival probability. For example, the parameters for the level
density influence the decay width very much~\cite{Zubov2009}.
Therefore, not only accurate nuclear properties such as the neutron
separation energy and the fission barrier, but also a proper form of
the level density are needed for an accurate prediction of the
survival probability of an excited superheavy nucleus.

\section{Summary~\label{sec:summary}}

In order to study the influence of the shell effects on the stability
of superheavy nuclei with excitation, the stability of excited
superheavy nuclei with $100 \leq Z \leq 134$ against neutron emission
and fission is studied by using a statistical model.

As an example, we present a systematic study of the survival
probability in the 1n-channel of these superheavy nuclei with $J=0$.
In this work, the properties of superheavy nuclei including the
neutron separation energies and shell correction energies are
consistently taken from predictions of the finite range droplet
model. The islands of stability of superheavy nuclei in their ground
state, e.g., the one around $Z=115$ and $N=179$, persist for excited
superheavy nuclei in the sense that the calculated survival
probabilities in the 1n-channel of excited superheavy nuclei on these
islands are maximized. This indicates that the survival probability
is mainly determined by the nuclear shell effects.

Finally it should be emphasized that the decay widths in different
channels and the survival probabilities of superheavy nuclei are very
sensitive to the nuclear properties such as the separation energy,
the ground state and saddle point deformations, the fission barriers,
and the level density, etc.. Therefore in the study of the decay
properties of superheavy nuclei, reliable predictions of these
properties from nuclear models is highly desirable.

\Acknowledgements{\bahao We thank G. G. Adamian, N. V.
Antonenko, A. Nasirov and A. S. Zubov for fruitful discussions
and suggestions. This work has been supported by the National
Natural Science Foundation of China (Grant Nos. 10705014,
10775012, 10875157, 10975100, and 10979066), Major State Basic
Research Development Program of China (Grant No. 2007CB815000),
and Knowledge Innovation Project of Chinese Academy of Sciences
(Grant Nos. KJCX2-EW-N01 and KJCX2-YW-N32). The computation of
this work was supported by Supercomputing Center, CNIC of CAS.
}

\normalsize \vskip0.3in\parskip=0mm \baselineskip 18pt
\renewcommand{\baselinestretch}{1.1}\footnotesize\parindent=4mm\bahao


\newpage

\end{document}